\documentclass[prb,12pt,nofootinbib,superscriptaddress,longbibliography]{revtex4-1}
\usepackage{graphicx}
\usepackage{hyperref}
\usepackage{amsmath}
\usepackage{amssymb}
\usepackage[usenames,dvipsnames]{color}

\usepackage{xcolor}
\usepackage{soul}

\begin{document}

\title{Electronic and magnetic characterization of epitaxial CrBr$_3$ monolayers}

\author{Shawulienu Kezilebieke \footnote[1]{These authors contributed equally.}}
\email{shawulienu.kezilebieke@aalto.fi}
\author{Orlando J. Silveira $^\mathrm{a}$}
\affiliation{Department of Applied Physics, Aalto University, FI-00076 Aalto, Finland}
\author{Md N. Huda $^\mathrm{a}$}
\affiliation{Department of Applied Physics, Aalto University, FI-00076 Aalto, Finland}
\author{Viliam Va\v{n}o}
\affiliation{Department of Applied Physics, Aalto University, FI-00076 Aalto, Finland}
\author{Markus Aapro}
\affiliation{Department of Applied Physics, Aalto University, FI-00076 Aalto, Finland}
\author{Somesh Chandra Ganguli}
\affiliation{Department of Applied Physics, Aalto University, FI-00076 Aalto, Finland}
\author{Jouko Lahtinen}
\affiliation{Department of Applied Physics, Aalto University, FI-00076 Aalto, Finland}
\author{Rhodri Mansell}
\affiliation{Department of Applied Physics, Aalto University, FI-00076 Aalto, Finland}
\author{Sebastiaan van Dijken}
\affiliation{Department of Applied Physics, Aalto University, FI-00076 Aalto, Finland}
\author{Adam S. Foster}
\affiliation{Department of Applied Physics, Aalto University, FI-00076 Aalto, Finland}
\affiliation{Nano Life Science Institute (WPI-NanoLSI), Kanazawa University, Kakuma-machi, Kanazawa 920-1192, Japan}

\author{Peter Liljeroth}
\affiliation{Department of Applied Physics, Aalto University, FI-00076 Aalto, Finland}


\begin{abstract}
The ability to imprint a given material property to another through proximity effect in layered two-dimensional materials has opened the way to the creation of designer materials. Here, we use molecular-beam epitaxy (MBE) for a direct synthesis of a superconductor-magnet hybrid heterostructure by combining superconducting niobium diselenide (NbSe$_2$) with the monolayer ferromagnetic chromium tribromide (CrBr$_3$). Using different characterization techniques and density-functional theory (DFT) calculations, we have confirmed that the CrBr$_3$ monolayer retains its ferromagnetic ordering with a magnetocrystalline anisotropy favoring an out-of-plane spin orientation. Low-temperature scanning tunneling microscopy (STM) measurements show a slight reduction of the superconducting gap of NbSe$_2$ and the formation of a vortex lattice on the CrBr$_3$ layer in experiments under an external magnetic field. Our results contribute to the broader framework of exploiting proximity effects  to realize novel phenomena in 2D heterostructures.
\end{abstract}

\maketitle

\newpage
Two-dimensional magnetic materials constitute an ideal platform to experimentally access the fundamental physics of magnetism in reduced dimensions \cite{Burch2018_review,Gibertini2019,Gongeaav4450}. Furthermore, because of the ease of fabricating  heterostructures, ferromagnetic van der Waals (vdW) materials present attractive opportunities for designer 2D magnetic \cite{Gong2017,Huang2017}, magnetoelectric \cite{Jiang2018}, and magneto-optical artificial heteromaterials \cite{Zhong2017}. As the different components in a vdW heterstructure only interact via weak vdW forces, the properties of the constituent materials are not strongly modified. This means that one can -- for example -- imprint magnetic properties of 2D magnets to the other layers without modifying their intrinsic properties and create novel spintronic and magnonic devices \cite{Lin2019,Wang2020,Ningrum2020}.

This designer concept can be utilized in systems combining magnetism with superconductivity to realize topological superconductivity \cite{Zhang2011_RMP,Sato2017}. It is currently attracting intense attention due to its potential role in building blocks for Majorana-based qubits for topological quantum computation \cite{Nayak2008,Sato2017,Lutchyn2018}. While there are very few potential real materials exhibiting topological superconductivity \cite{Zhang2018_Science,Wang2018_Science,Zhu2020_science,Wang2020_Science}, in a designer material the desired physics emerges from the engineered interactions between the different components. For topological superconductivity, one needs to combine s-wave superconductivity with magnetism and spin-orbit coupling to create an artificial topological superconductor \cite{Buzdin2005,Sato2017}. However, the coupling between the components is highly sensitive to the interfacial structure and electronic properties \cite{Zhong2020,Gibertini2019} and thus, vdW materials with atomically sharp and highly uniform interfaces are an attractive platform with which to realize and harness exotic electronic phases arising in designer materials. 

Layered materials  that  remain  magnetic  down  to  the  monolayer  (ML)  limit  have  been  recently demonstrated \cite{Gong2017,Huang2017,Huang2020}. While the first reports relied on mechanical exfoliation for the sample preparation, related materials CrBr$_3$ and Fe$_3$GeTe$_2$ have also been grown using molecular-beam epitaxy (MBE) in ultra-high vacuum (UHV) \cite{Chen2019,Liu2017}, which is essential for realizing clean edges and interfaces. The inherent lack of surface bonding sites due to the layered nature of these materials prevents chemical bonding between the layers and results in a better control of the interfaces. Recently, we have successfully fabricated a superconducting ferromagnetic hybrid system based on vdW heterostructures using MBE \cite{Kezilebieke2020a,Kezilebieke2020}. More importantly, by combining spin-orbit coupling, 2D ferromagnetic CrBr$_3$, and superconducting niobium diselenide (NbSe$_2$), we have demonstrated the existence of the one-dimensional Majorana edge modes using low-temperature scanning tunneling  microscopy (STM) and spectroscopy (STS) \cite{Kezilebieke2020}. However, for future applications, further systematic studies are desired for a better understanding of the electronic and magnetic properties of the 2D ferromagnetic CrBr$_3$ / NbSe$_2$ heterostructures.

In this work, we report a detailed characterisation of the CrBr$_3$/NbSe$_2$ hybrid system using low-temperature STM/STS and X-ray photo-electron spectroscopy (XPS). Combining magneto-optical Kerr effect (MOKE) measurements and density functional calculations (DFT), we unambiguously confirm the ferromagnetism of the CrBr$_3$ islands on NbSe$_2$ substrate. Our results give further experimental information on the magnetic properties of CrBr$_3$, and demonstrate a clean and controllable platform for creating superconducting-magnetic hybrid systems with a great potential for integration into future electronic devices that could be controlled externally through electrical \cite{Jiang2018}, mechanical \cite{Wu2019}, chemical \cite{Jiang2018a}, or optical means \cite{Zhang2019}.

\begin{figure}[!h]
	\centering
		\includegraphics[width=0.9\textwidth]{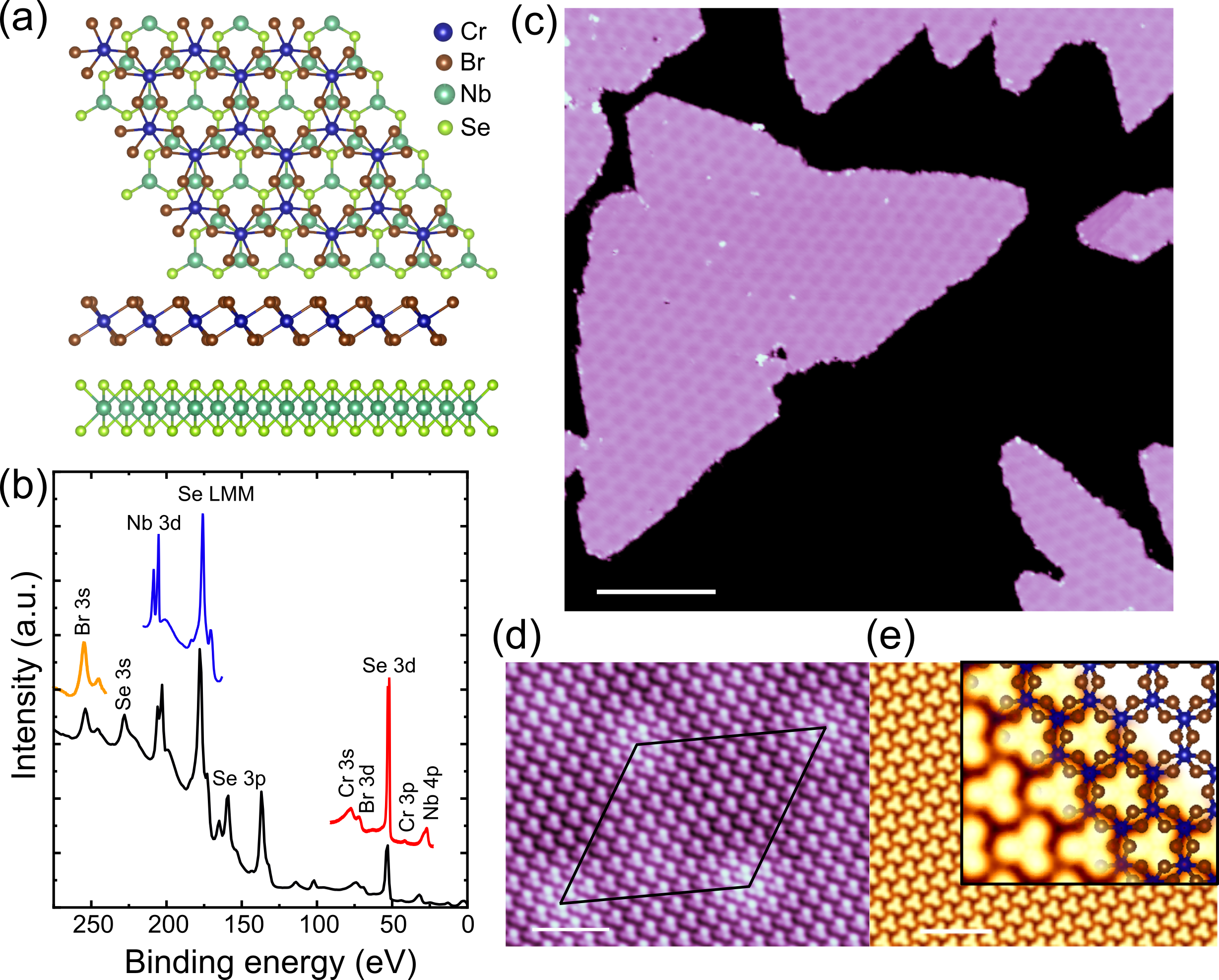}
	\caption{Growth of CrBr$_3$ on NbSe$_2$. (a) Computed top and side views of the CrBr$_3$-NbSe$_2$ heterostructure. (b) Core-level spectra of CrBr$_3$-NbSe$_2$ heterostructure. The pass energy was 80 eV for the black curve and 20 eV for colored curves (c) STM image of a monolayer thick CrBr$_3$ island grown on NbSe$_2$ using MBE (STM feedback parameters: $V_\mathrm{bias} = +1$ V, $I$ = 10 pA, scale bar: 39 nm). (d)  Atomically resolved image on the CrBr$_3$ layer. The moiré unit cell is denoted by the large rhombus (STM feedback parameters: $V_\mathrm{bias} = +2$ V, $I = 0.5$ nA, scale bar: 3 nm). (e) Calculated constant-current STM image of a monolayer of CrBr$_3$ on NbSe$_2$: $V_\mathrm{bias} = +0.5$ V, scale bar: 3 nm. The inset in (e) shows a zoom in the calculated STM images, with the atomic structure of the CrBr$_3$ superimposed.}
	\label{fig1}	
\end{figure}

Figure~\ref{fig1}a shows the top and side views of the most stable CrBr$_3$-NbSe$_2$ heterostructure geometry obtained by density functional theory (DFT) calculations (details are given in the Methods section), where the single-layer CrBr$_3$ consists of Cr atoms that are arranged in a honeycomb lattice structure and each atom is surrounded by an octahedron of six Br atoms. The energetically most favourable stacking has one Cr atom located on top of a Se$_2$ pair, while the other Cr atom is on top of the hole site of the NbSe$_2$. The fully relaxed lattice parameters of the heterostructures reveal that the CrBr$_3$ is strained by about 7 \%, while the NbSe$_2$ is compressed by less than 2 \%, which are not sufficient to drastically affect their electronic and magnetic properties \cite{Webster2018,Xu2014} (see Supporting Information). X-ray photoelectron spectroscopy (XPS) (Figure~\ref{fig1}b) displays sharp characteristic peaks for Cr and Br, demonstrating the purity of the CrBr$_3$ film as well as consistency with previous reports on bulk samples \cite{Carver2020,Pollini2003}. Figure~\ref{fig1}c shows a large scale STM image illustrating the typical morphology of the single-layer CrBr$_3$ films grown on NbSe$_2$. The CrBr$_3$ islands are atomically flat and up to 200 nm in size, while clean areas of the bare NbSe$_2$ substrate are clearly exposed. Figure~\ref{fig1}d shows an atomically resolved STM image of the CrBr$_3$ monolayer, revealing periodically spaced triangular protrusions. These features are formed by the three neighbouring Br atoms as highlighted in the Fig.~\ref{fig1}e showing the simulated STM topographic image obtained through DFT calculations. Finally, the STM topography exhibits a clear, well-ordered superstructure with 6.3 nm periodicity. This superstructure is explained by the fact that when the CrBr$_3$ and NbSe$_2$ lattices are overlaid, 19 NbSe$_2$ unit cells accommodate 10 unit cells of CrBr$_3$, thus forming a  6.3 nm $\times$ 6.3 nm superstructure (a moir\'e pattern) as observed in the STM images of CrBr$_3$ as shown in  Figure~\ref{fig1}d. The rotationally misaligned CrBr$_3$ domains have a slight effect the on moiré periodicity (supporting info).

\begin{figure}[!h]
	\centering
		\includegraphics [width=0.9\textwidth]{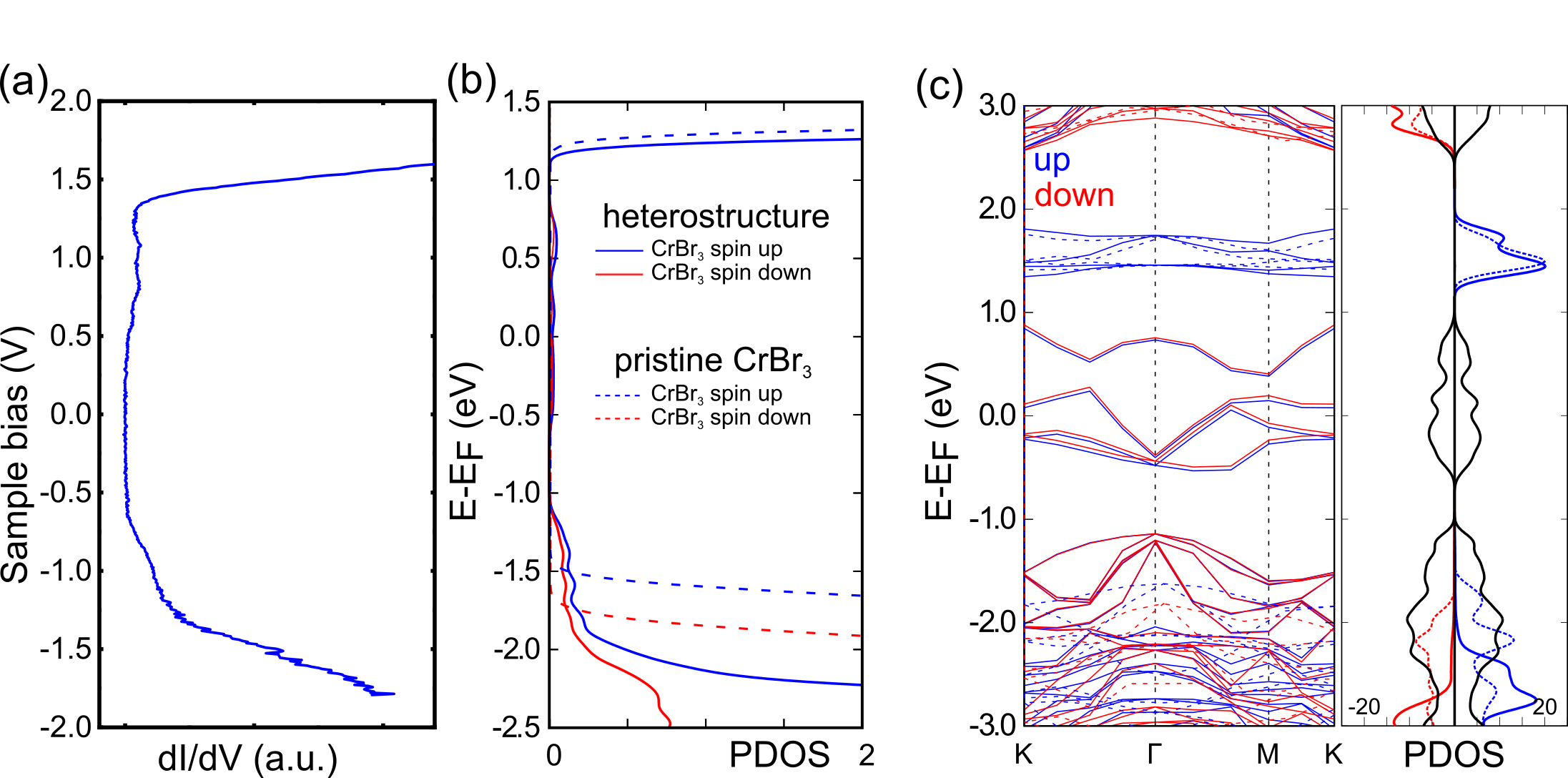}
	\caption{(a) Typical long-range experimental d$I$/d$V$ spectra on a ML CrBr$_3$ on the NbSe$_2$ substrate. (b) Simulated spin polarized projected density of states (PDOS) on the CrBr$_3$ layer of the CrBr$_3$-NbSe$_2$ heterostructure (continuous lines) and on pristine CrBr$_3$ layer (dashed lines). (c) Band structure and PDOS of the pristine CrBr$_3$ layer (dashed lines) and CrBr$_3$-NbSe$_2$ heterostructure (continuous lines). The black continuous line in (c) is the PDOS on the NbSe$_2$ substrate in the CrBr$_3$-NbSe$_2$ heterostructure.}
	\label{fig2}	
\end{figure}

\begin{figure}[!th]
	\centering
		\includegraphics [width=0.9\textwidth]{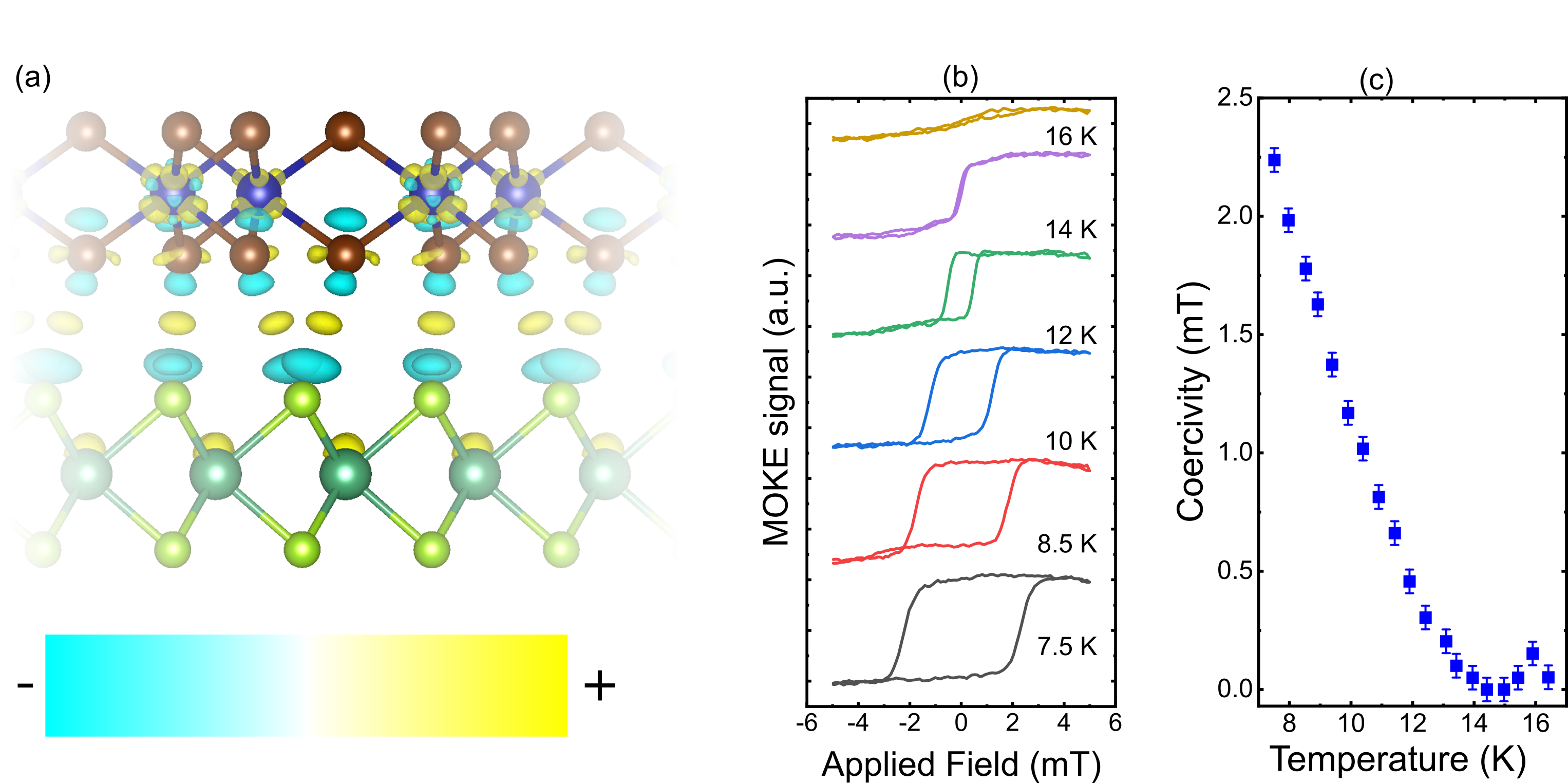}
	\caption{(a) Differential charge density of the CrBr$_3$-NbSe$_2$ heterostructure, where the yellow and blue colors indicate charge accumulation (+) and depletion (-), respectively. (b) Magnetic hysteresis in monolayer CrBr$_3$ on NbSe$_2$ at several different temperatures (indicated in the figure). (c) The temperature dependence of the coercive field of CrBr$_3$ on NbSe$_2$. The coercive field decreases as the temperature increases until it vanishes at 16 K.}
	\label{fig3}	
\end{figure}

We experimentally determined the electronic structure of single-layer CrBr$_3$-NbSe$_2$ using scanning tunneling spectroscopy (STS) measurements and compared the results directly with DFT calculations. Figure~\ref{fig2}a shows a typical STM d$I$/d$V_\mathrm{b}$ spectrum of single-layer CrBr$_3$ acquired over a large bias range. In the filled state regime (bias voltage $V_\mathrm{b}<0$), the spectrum is relatively flat and featureless until $V_\mathrm{b}\sim -1$ V is reached, where the d$I$/d$V_\mathrm{b}$ slightly increases, and a steep rise is seen at $V_\mathrm{b}\sim$ -1.5 V. The dominant feature in the empty state regime ($V_\mathrm{b}>0$) of the d$I$/d$V_\mathrm{b}$ spectrum starts with a small bump around $\sim 0.5$ V and a steep rise in d$I$/d$V_\mathrm{b}$ at $\sim 1.2$ V, as shown in Figure~\ref{fig2}a. Those features, apart from a hard shift of $\sim 0.5$ eV, are consistent with the DFT spin polarized projected density of states (PDOS) on the CrBr$_3$ layer of the heterostructure shown in Figure~\ref{fig2}b. The steep rises in the d$I$/d$V_\mathrm{b}$ spectrum are around 3 V apart and mostly arise from the spin up bands of the CrBr$_3$ layer: its electronic properties are well preserved in the heterostructure as compared to the pristine CrBr$_3$ layer, as can be seen in Figure~\ref{fig2}b and c. The band structures and PDOS shown in Figure~\ref{fig2}c reveal that the bands in the [-2.0,-1.0] eV and [-1.0,1.0] eV windows have a majority contribution from the NbSe$_2$ layer, where the NbSe$_2$'s $d$-band in the [-1.0,1.0] eV window is completely preserved and slightly spin polarized due to the proximity with the magnetic CrBr$_3$. However, the PDOS on the CrBr$_3$ layer in the [-2.0,-1.0] eV and [-1.0,1.0] eV windows are non-zero and consistent with the bumps in the d$I$/d$V_\mathrm{b}$ spectrum, which we attribute to charge reconfiguration in the heterostructure as shown in Figure~\ref{fig3}a.

After the electronic characterization of the samples, we will next focus on their magnetic properties. The isolated CrBr$_3$ layer has a predicted out-of-plane magnetic moment of 6.000 $\mu_B$ per unit cell, where each Cr atom has three unpaired electrons. Our DFT calculations show that the magnetism of the CrBr$_3$ layer is well-preserved in the heterostructure, which shows a slightly larger magnetic moment of 6.097 $\mu_B$ per unit cell due to induced magnetization on the NbSe$_2$ layer. 
The PDOS on the CrBr$_3$ layer of the heterostructure reveals that the majority spin up channel has a band gap of 3 eV, while the minority spin down channel has a band gap of around 5 eV, both shown in blue and red, respectively, in Figure \ref{fig2}c.

We study the truly 2D itinerant ferromagnetism in CrBr$_3$ monolayer on NbSe$_2$ using magneto-optical Kerr effect (MOKE) microscopy (details of the experimental procedures are given the Methods section). The magnetic field was applied perpendicular to the sample. Figure~\ref{fig3}b shows the MOKE signal of a CrBr$_3$ monolayer on NbSe$_2$ as a function of the external magnetic field at several different temperatures. Notably, monolayer CrBr$_3$ is ferromagnetic, as evidenced by the prominent hysteresis seen here. MOKE measurements further reveal that as the temperature is increased, the hysteresis loop shrinks and eventually disappears at a transition temperature $T_\mathrm{c}$ of about 16 K. The $T_\mathrm{c}$ can also be extracted from a temperature dependence of the coercivity $H_\mathrm{c}$ (blue squares in Figure~\ref{fig3}c), where the onset of $H_\mathrm{c}$ in the CrBr$_3$ monolayer clearly occurs at around 16 K. The same behaviour was observed for mechanically exfoliated monolayer CrBr$_3$ flake, where the transition temperature $T_\mathrm{c}$ was around 20 K \cite{Kim2019}. Moreover, the out-of-plane coercive field $H_\mathrm{c}$ for exfoliated monolayer CrBr$_3$ flake is 4 mT \cite{Kim2019, Zhang2019,Kim2019a} while in our MBE grown monolayer CrBr$_3$ on NbSe$_2$ substrate the out-of-plane coercive field increases up to $\sim 2.5$ mT at the lowest temperatures we can reach in our MOKE setup as shown in Figure~\ref{fig3}c. 

\begin{figure}[!h]
	\centering
		\includegraphics [width=0.9\textwidth]{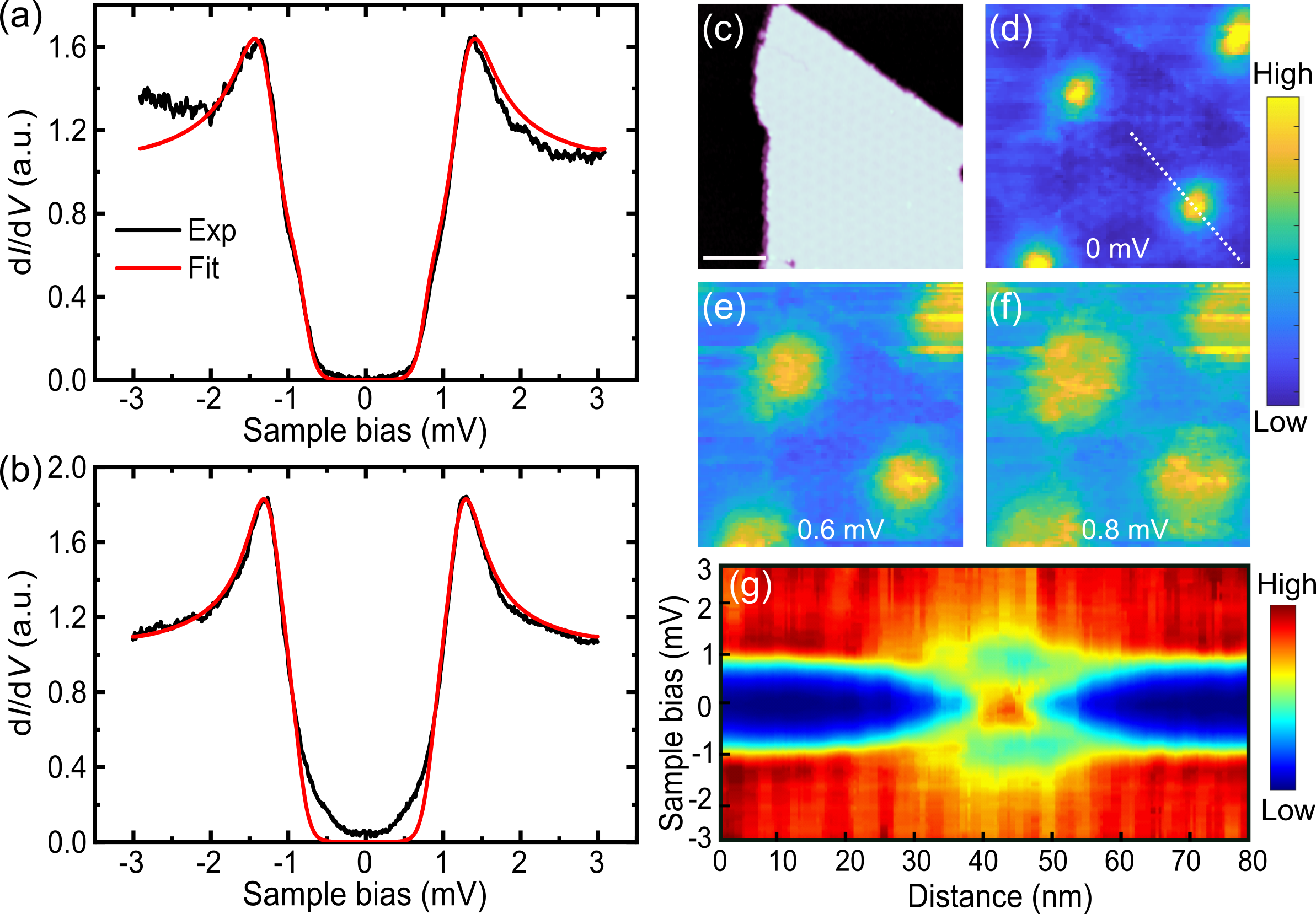}
	\caption{(a,b) Experimental d$I$/d$V$ spectroscopy (black) on the NbSe$_2$ substrate (a) and in the middle of a CrBr$_3$ island (b) measured at $T=350$ mK. We also show a fit to a double gap s-wave BCS-type model (red). (c) STM topography image (STM feedback parameters: $V_\mathrm{bias} = +1$ V, $I$ = 10 pA, scale bar: 22 nm). (d-f) Vortex imaging on CrBr$_3$/NbSe$_2$ heterostructure. We have recorded grid spectroscopy ($100 \times 100$ spectra) over an area of $110 \times 110$ nm$^2$ at $T= 350$ mK under an applied out-of-plane magnetic field of 0.65 T. The maps are obtained from full spectroscopic scans from -3 mV to 3 mV at each pixel at the indicated bias voltages. (g) Line scan of the d$I$/d$V_b$ signal along the white line marked in panel (d).}
	\label{fig4}	
\end{figure}

After confirming the existence of ferromagnetism on the monolayer CrBr$_3$ on NbSe$_2$ substrate, it is interesting to see the effects of the superconducting substrate NbSe$_2$ on the monolayer CrBr$_3$. Isolated CrBr$_3$ is a ferromagnetic insulator; however, due to the monolayer thickness, it is possible to tunnel through such a structure with STM. In addition, due to the charge transfer at the interface between CrBr$_3$ and NbSe$_2$, the heterostructure itself has a metallic nature. Therefore, one can expect a measurable interaction between CrBr$_3$ and NbSe$_2$. The superconductivity of CrBr$_3$/NbSe$_2$ heterostructure was studied by STM at $T=350$ mK. Figures~\ref{fig4}a,b show experimental d$I$/d$V_\mathrm{b}$ spectra (raw data) taken on a bare NbSe$_2$ and on monolayer CrBr$_3$, respectively. The d$I$/d$V_\mathrm{b}$ spectrum of bare NbSe$_2$ (Figure ~\ref{fig4}a) has a hard gap with an extended region of zero differential conductance around the Fermi energy. d$I$/d$V_\mathrm{b}$ spectra were fitted by the McMillan two-band model \cite{Noat2015}, with parameters $\Delta_1 = 1.28$ meV,$\gamma_1 = 0.38$ and $\Delta_2 = 0.74$ meV,$\gamma_1 = 0.13$ meV, respectively. In contrast, the spectra taken in the middle of the CrBr$_3$ island have small but distinctly non-zero differential conductance inside the gap of the NbSe$_2$ substrate. We observe pairs of conductance onsets at $\pm 0.3$ mV around zero bias. This feature results from the formation of Shiba-bands in the NbSe$_2$ caused by the induced magnetization from the CrBr$_3$ island \cite{Kezilebieke2020}. These extra features are not reproduced by the two-band model (Figure~\ref{fig4}b). Nevertheless, a double gap s-wave BCS-type fitting gives a slight reduction of both gap parameters ($\Delta_1= 1.20$ meV, $\gamma_1 = 0.40$ meV and $\Delta_2 = 0.73$ meV, $\gamma_2 = 0.50$ meV respectively).

To obtain more detailed insight into the superconductivity on CrBr$_3$/NbSe$_2$ heterostructure, we investigate the dependence of our d$I$/d$V_\mathrm{b}$ spectra under an applied out-of-plane magnetic field. In a type-II SC such as NbSe$_2$, we would expect to observe an Abrikosov vortex lattice in a d$I$/d$V_\mathrm{b}$ map acquired near the energy of the superconducting gap \cite{Hess1989}. Figs.~\ref{fig4}d-f show d$I$/d$V_\mathrm{b}$ grid maps at 0, 0.6 and 0.8 mV bias voltages, respectively. The maps are recorded under 0.65 T out-of-plane magnetic field on an area with both CrBr$_3$ islands and bare NbSe$_2$ surface (Figure~\ref{fig4}c). It is seen from Figures~\ref{fig4}d-f that the vortices exhibit a highly ordered hexagonal lattice similar to those observed on the clean NbSe$_2$ surface. This is the first time vortices have been clearly observed in a hybrid ferromagnet-superconductor-system. We measured the spatial variation of the d$I$/d$V_\mathrm{b}$ spectra as a function of distance away from the vortex center (along the dashed line in Figure~\ref{fig4}d). The results are given in Figure~\ref{fig4}g, which shows the measured d$I$/d$V_\mathrm{b}$ as functions of distance and sample bias V on a color scale. One can see that only one peak appears at zero-bias in the d$I$/d$V_\mathrm{b}$ spectra near the vortex center, and this peak splits into two away from the vortex core. The splitting energy increases linearly with distance. One of the intriguing properties of a topological superconductor is that vortices on its surface are expected to host Majorana zero modes \cite{Sun2017}. This mode results in a peak in the local density-of-states at the Fermi energy in the center of the vortex. In contrast to bound states in vortices on conventional superconductors, a Majorana mode should not split in energy away from the vortex center\cite{Sun2017,Xu2015,Yin2015}. Due to the broadening of the resonance in the vortex spectra, we cannot resolve the individual components in the d$I$/d$V_\mathrm{b}$ spectra and the zero bias feature persists up to 5 nm from the vortex core before the clearly split features can be observed. This is within the range of spatial distributions of the Majorana zero mode reported in the literature (a few nm up to 30 nm) \cite{,Liu2018,Xu2015,Yin2015,Wang2018}. In addition, it is important to note that Majorana zero modes in vortex cores have only been observed at low magnetic field (e.g.~0.1 T for Bi$_2$Se$_3$/NbSe$_2$ \cite{Xu2015,Wang2018} and its amplitude is expected to decay exponentially as the external field is increased. In order to confirm whether the vortices in our system host Majorana zero modes at their cores, further experiments, in particular using spin-polarized tunneling measurements \cite{Sun2016}, are clearly necessary.\\

In summary, we have provided experimental evidence for the realisation of a 2D ferromagnet-superconductor van der Waals heterostructure. More importantly, we have experimentally confirmed that the CrBr$_3$ monolayer retains its ferromagnetic ordering with a magnetocrystalline anisotropy favoring an out-of-plane spin orientation on NbSe$_2$. Our DFT calculations showing an induced moment in Nb from hybridization with Cr $d$-orbitals confirm the imprinting of magnetic order on NbSe$_2$ from a 2D vdW magnetic insulator. Our results provide a broader framework for employing other proximity effects to tailor materials and realize novel phenomena in 2D heterostructures. 

\section*{Methods}
\emph{Molecular‐beam epitaxy (MBE) sample growth:} The CrBr$_3$ thin film was grown on a freshly cleaved NbSe$_2$ substrate by compound source MBE. The anhydrous CrBr$_3$ powder of 99 \% purity was evaporated from a Knudsen cell. Before growth, the cell was degassed up to the growth temperature $350^\circ$C until the vacuum was better than $1\times10^{-8}$ mbar. The sample was heated by electron beam bombardment and temperatures were measured using an optical pyrometer. The growth speed was determined by checking the coverage of the as-grown samples by STM. The optimal substrate temperature for the growth of CrBr$_3$ monolayer films was $\sim270^\circ$C. Below this temperature, CrBr$_3$ forms disordered clusters on the NbSe$_2$ surface. The NbSe$_2$ crystal is directly mounted on a sample holder using a two component conducting silver epoxy which only allows us to heat the sample up to $\sim300^\circ$C.

\emph{Scanning tunneling microscopy (STM) and spectroscopy (STS) measurements:} After the sample preparation, it was inserted into a low-temperature STM (Unisoku USM-1300) housed in the same UHV system and all subsequent experiments were performed at $T = 350$ mK. STM images were taken in constant-current mode. d$I$/d$V_\mathrm{b}$ spectra were recorded by standard lock-in detection while sweeping the sample bias in an open feedback loop configuration, with a peak-to-peak bias modulation of 30-50 $\mu$V for a small bias range and 10 mV for a lager bias range, respectively at a frequency of 707 Hz. Spectra from grid spectroscopy experiments were normalized by the normal state conductance, i.e.~d$I$/d$V_\mathrm{b}$ at a bias voltage corresponding to a few times the superconducting gap.\\
\emph{Sample transfer:} After the sample growth, the sample was transferred to the load lock ($\sim10^{-9}$ mbar) and then to the directly connected glove bag. The load lock is slowly vented  with pure nitrogen to the glove bag atmosphere, and the sample is then transferred into the glove bag with a magnetic transfer rod for XPS and MOKE measurements.

\emph{X-Ray photoelectron spectroscopy (XPS) experiments:} The XPS spectra were measured using a Kratos Axis Ultra system, equipped with monochromatic Al K$_\alpha$ X-ray source. All measurements were performed using an analysis area of $0.3 \rm{mm} \times 0.7 \rm{mm}$. The black curve in Fig 1b was measured using 80 eV pass energy and 1 eV energy step whereas the colored curves were taken with 20 eV pass energy and 0.1 eV energy step. The energy calibration was done using the C 1s peak at 284.8 eV. 

\emph{Magneto-optical Kerr effect (MOKE) measurements:}
MOKE was carried out using an Evico Magnetics system based on a Zeiss Axio Imager D1 microscope with a Hamamatsu C4742-95 digital camera. The sample was placed in a Janis research ST-500 cold finger cryostat with optical access and cooled using liquid He. Imaging of the sample was carried out in a polar Kerr configuration. Hysteresis loops were constructed from images taken using a long working distance 100x lens after background image subtraction and corrected for a linear background slope due to the Faraday effect on the lens.

\emph{DFT calculations:} Calculations were performed with the DFT methodology as implemented in the periodic plane-wave basis VASP code \cite{PhysRevB.54.11169,KRESSE199615}. Atomic positions and lattice parameters were obtained by fully relaxing all structures using the spin-polarized Perdew-Burke-Ernzehof (PBE) functional \cite{PhysRevLett.77.3865} including Grimme's semiempirical DFT-D3 scheme for dispersion correction \cite{doi:10.1063/1.3382344}, which is important to describe the van der Waals (vdW) interactions between the CrBr$_3$ and the NbSe$_2$ layers. The stacking of the layers used in our calculations is the same used in a previous work \cite{Kezilebieke2020}. The convergence criterion of self-consistent field (SCF) computation was set to 10$^{-5}$ eV and the threshold for the largest force acting on the atoms was set to less than 0.012 eV/\AA{}. A vacuum layer of 12 \AA{} was added to avoid mirror interactions between periodic images. Further calculations of band structures and density of states were realized using the hybrid Heyd-Scuseria-Ernzerhof (HSE06) functional \cite{doi:10.1063/1.1760074, doi:10.1063/1.2204597, doi:10.1063/1.1564060}, which improves the description of the band structure as compared to the PBE functional. The interactions between electrons and ions were described by PAW pseudopotentials, where $4s$ and $4p$ shells were added explicitly as semicore states for Nb and $3p$ shells for Cr. An energy cutoff of 550 eV is used to expand the wave functions and a systematic $k$-point convergence was checked, where the total energy was converged to the order of 10$^{-4}$ eV.  Spin polarization was considered by setting an initial out-of-plane magnetization of 3 $\mu_B$ per Cr atom and zero otherwise.

\section*{Acknowledgments}
This research made use of the Aalto Nanomicroscopy Center (Aalto NMC) facilities and was supported by the European Research Council (ERC-2017-AdG no.~788185 ``Artificial Designer Materials'') and Academy of Finland (Academy professor funding no.~318995 and 320555, Academy postdoctoral researcher no.~309975). Computing resources from the Aalto Science-IT project and CSC, Helsinki are gratefully acknowledged. ASF has been supported by the World Premier International Research Center Initiative (WPI), MEXT, Japan.

\bibliography{CrBr}

\end{document}